\documentclass[aps,prb,amsmath,amssymb,twocolumn,10pt]{revtex4-1}

\usepackage{times}
\usepackage{graphicx}
\usepackage{hyperref}

\usepackage{dcolumn}% Align table columns on decimal point
\usepackage{multirow}
\usepackage{bm}% bold math

\begin{document}
\title{Role of oxygen vacancy in the spin-state change and magnetic ordering in SrCoO$_{3-\delta}$}

\author{Jinyoung Lim}
\author{Jaejun Yu}
\email{jyu@snu.ac.kr}
\affiliation{Center for Theoretical Physics, Department of Physics and Astronomy, Seoul National University, Seoul 08826, Korea}

\date{\today}

\begin{abstract}
  We present the first-principles investigation of the structural, electronic, and magnetic properties of SrCoO$_{3-\delta}$ ($\delta=0, 0.25, 0.5$) to understand the multivalent nature of Co ions in SrCoO$_{3-\delta}$ along the line of topotactic transition between perovskite SrCoO$_{3}$ and brownmillerite SrCoO$_{2.5}$. From the on-site Coulomb interaction $U$-dependent ground state of stoichiometric SrCoO$_{3}$, we show the proximity of its metallic ferromagnetic ground state to other antiferromagnetic states. The structural and magnetic properties of SrCoO$_{3-\delta}$ depending on their oxygen-content provide an interesting insight into the relationship between the Co-Co distances and the magnetic couplings so that the spin-state transition of Co spins can understood by the change of $pd$-hybridization depending on the Co-Co distances. The \emph{strong} suppression of the $dp\sigma$-hybridization between Co $d$ and O $p$ orbitals in brownmillerite SrCoO$_{2.5}$ brings on the high-spin state of Co$^{3+}$ $d^{6}$ and is responsible for the antiferromagnetically ordered insulating ground state. The increase of effective Co-Co distances driven by the presence of oxygen vacancies in SrCoO$_{3-\delta}$ is consistent with the reduction of the effective $pd$-hybridization between Co $d$ and O $p$ orbitals. We conclude that the configuration of neighboring Co spins is shown to be crucial to their local electronic structure near the metal-to-insulator transition along the line of the topotactic transition in SrCoO$_{3-\delta}$. Incidentally, we also find that the \textit{I2mb} symmetry of SrCoO$_{2.5}$ is energetically stable and exhibits ferroelectricity via the ordering of CoO$_{4}$ tetrahedra, where this polar lattice can be stabilized by the presence of a large activation barrier.
\end{abstract}

%---------------------------------------------
% \pacs{75.10.-b, 61.72.-y,75.30.Et,77.84.-s}
% Disciplines - Condensed Matter & Materials Physics
% Research Areas - Metal-Insulator Transitions, topotactic, Vacancies,
% Physical Systems - Functional materials, Oxides, Perovskite, Thin films
% Techniques - DFT+U
%-----------------------------------------------

\maketitle

\section{Introduction}
\label{sec:introd}

Transition metal oxides have received a lot of interests for their fascinating physical properties such as superconductivity, magnetism, and ferroelectricity. Such features are often associated with phase transitions driven by correlation effects arising from electron-electron interactions.\cite{Imada:1998aa} Near the transition, the interplay among the spin, charge, and orbital degrees of freedom is so crucial that a small change in doping, strain, or temperature can develop a system into different orderings, for example, ferroelectric, ferromagnetic, or even orbital and charge orderings in multiferroic materials.\cite{Spaldin:2005aa} These orderings can be tuned by external electric, magnetic, or stress field, and the cross-couplings between them enable critical multifunctional properties, which makes these transition metal oxides as a technologically significant class of materials.

Recently, a transformation between two distinct topotactic phases of the brownmillerite SrCoO$_{2.5}$ (BM-SCO)\cite{Takeda:1972ab} and the perovskite SrCoO$_{3}$ (P-SCO)\cite{Long:2011aa} was reported to show a novel oxygen-content-dependent phase transition.\cite{Jeen:2013aa,Jeen:2013ab} It was also demonstrated that the control of the contents of oxygen vacancies by epitaxial strain and temperature in thin films can be used to tune their electronic and magnetic properties.\cite{Callori:2015aa,Petrie:2016aa,Copie:2017aa} These cobalt-based oxides were suggested to be candidates for various technical applications such as solid oxide fuel cells,\cite{Shao:2004aa} catalysts,\cite{Gangopadhayay:2009aa} oxygen membranes,\cite{Vashook:1999aa} and resistive RAM (random access memory) devices.\cite{Tambunan:2014aa}

It is now well known that oxygen stoichiometry in SrCoO$_{3-\delta}$ plays a crucial role in determining their structural, electronic, and magnetic properties including metal-to-insulator and ferromagnetic-to-antiferromagnetic transitions. An optical spectroscopy study combined with first-principles calculations by Choi \textit{et al.}\cite{Choi:2013aa} showed that SrCoO$_{3-\delta}$ ($0\le\delta\le 0.5$) exhibits a reversible lattice and electronic structure evolution according to the change of oxygen stoichiometry. Their interpretation of the metal-to-insulator transition was based on the two stable electronic configurations in P-SCO and BM-SCO: Co$^{4+}$ (3$d^{5}$) in a ferromagnetic (FM) metallic state for P-SCO ($\delta=0$) and Co$^{3+}$ (3$d^{6}$) in an antiferromagnetic (AFM) insulating state for BM-SCO ($\delta=0.5$). The formation of CoO$_{4}$ tetrahedral layers, characterized by one-dimensionally ordered chains of oxygen vacancies,\cite{Abakumov:2005ab,DHondt:2008aa,Munoz:2008aa} in BM-SCO was identified as a key structural feature to disrupt the double exchange leading to an insulating state. Further, the spectroscopic evidence for the split $t_\mathrm{2g}$ bands in the tetrahedral layer was shown to be consistent with the calculated electronic structure for the tetrahedral layers in BM-SCO.

The modification of Co valence states may have been possible due to the presence of two structurally distinct topotactic phases of P-SCO and BM-SCO. Perhaps, additional oxygens in the oxygen vacancy channels in BM-SCO can adapt the valence state of Co to change. However, it is not clear how the multivalent nature of Co ions are attributed to the metal-to-insulator and FM-to-AFM transitions in SrCoO$_{3-\delta}$ ($0\le\delta\le 0.5$), especially, in terms of its oxygen-content dependence in SrCoO$_{3}$. Recent first-principles calculations for P-SCO\cite{Lee:2011aa} showed that the ground state is an intermediate spin state of Co$^{4+}$, which in good agreement with experiments.\cite{Long:2011aa,Bezdicka:1993aa} Then,
they predicted that the epitaxial strain on P-SCO could induce electronic and magnetic phase transitions from FM-metal to AFM-insulator with ferroelectricity for both compressive and tensile strains.\cite{Lee:2011aa} It implies that the ground state of P-SCO is close to either FM-metal or AFM-insulating ground states even without changing the Co valence state.

The spin state of Co ions in the pristine P-SCO has been a subject of debates in connection with the low-spin (LS) to high-spin (HS) transition of LaCoO$_{3}$.\cite{Thornton:1982aa,Rao:2004aa}  An earlier atomic multiplet calculation suggested that the intermediate-spin (IS) ground state is possible for Co$^{4+}$ where the $d^{6}\underbar{L}$ state dominates the ground state and, consequently, the hole residing in the oxygen ligand, which is antiferromagnetically coupled to neighboring Co ions, becomes itinerant and couples the high-spin Co $d^{6}$ ions ferromagnetically.\cite{Potze:1995aa} The IS state of Co ions depends on the competition between the crystal field strength and Hund’s coupling \cite{Zhuang:1998aa} and the presence of oxygen vacancy.\cite{Hoffmann:2015aa} Therefore, the spin state of Co ions in P-SCO or SrCoO$_{3-\delta}$ ($0\le\delta\le 0.5$) can be in proximity to HS, LS, or even IS states.

Here, we investigate the structural, electronic, and magnetic properties of SrCoO$_{3-\delta}$ ($\delta=0, 0.25, 0.5$)
to understand the multivalent nature of Co ions in SrCoO$_{3-\delta}$ ($0\le\delta\le 0.5$). We carried out first-principles calculations by using density-functional theory within the GGA+$U$ method, as described in Sec.~\ref{sec:methods}. We calculate the $U_{\mathrm{eff}}$-dependent magnetic structures to address the issues of the proximity of the FM-metallic P-SCO to other AFM states. The results of structural and magnetic properties for perovskite SrCoO$_{3-\delta}$ in Sec.~\ref{sec:p-sco} provide an interesting insight into the relation between the Co-Co distances and the magnetic couplings, thereby explaining of the change of the Co spin state in terms of the strength of on-site Coulomb interactions and oxygen vacancies. In Sec.~\ref{sec:brownmillerite-sco}, we present the calculation results for the structural, electronic and magnetic properties of brownmillerite SrCoO$_{2.5}$. The \emph{strong} suppression of the $dp\sigma$-hybridization between Co $d$ and O $p$ orbitals brings on the HS state of Co$^{3+}$ d$^{6}$ and the antiferromagnetically ordered state. The reduction of the effective $pd$-hybridization is consistent with the increase of effective Co-Co distances in BM-SCO, which in turn affect the Co spin state. It is interesting to note that the BM-SCO with the \textit{I2mb} structure, which is found to be energetically the most stable, exhibits ferroelectricity via the ordering of CoO$_{4}$ tetrahedra and the polar structure of \textit{I2mb} may be stabilized by the large activation barrier among different structural configurations. We also examine the structural, electronic and magnetic properties of SrCoO$_{2.75}$ with an intermediate oxygen content between P-SCO and BM-SCO in Sec.~\ref{sec:intermediate-structure}. We consider a stacking of perovskite and brownmillerite layers as one of the possible structural configurations for SrCoO$_{2.75}$, which is close to the metal-to-insulator boundary between FM-metallic P-SCO and G-type antiferromagnetic (G-AFM) insulating BM-SCO. It turns out that the local spin configuration is crucial to the electronic structure properties and the overall electronic property can be complicated depending on the spin configurations. In Sec.~\ref{sec: discussion}, we summarize the effect of $pd$-hybridization on the magnetic ordering as well as the spin state of Co ions and, consequently, the electronic properties of SrCoO$_{3-\delta}$, which depend on the contents of oxygen vacancy.

\section{Methods}
\label{sec:methods}

The first-principles calculations were performed by using density-functional theory (DFT) within the generalized gradient approximation GGA+$U$ method. The projector augmented wave (PAW)\cite{Kresse:1999aa} pseudopotentials are adopted as implemented in the Vienna \textit{ab initio} simulation package (VASP) code.\cite{Kresse:1996aa} The wave functions are expanded in a plane waves basis with an energy cutoff of 600 eV. For SrCoO$_3$, a $\sqrt{2}\times\sqrt{2}\times 2$ unit cell are used to accommodate G-type antiferromagnetic ordering with $8\times 8\times 6$ Monkhorst-Pack $\mathbf{k}$-point grids. For the brownmillerite SrCoO$_{2.5}$, we use a supercell corresponding to $\sqrt{2}\times\sqrt{2}\times 4$ perovskite cells with $6\times 6\times 2$ $\mathbf{k}$-point grids.

The lattice structures are relaxed until the residual forces converge within 0.001 eV/\AA\  (SrCoO$_3$) or 0.01 eV/\AA\ (SrCoO$_{2.5}$). We use the effective on-site Coulomb interaction $U_\mathrm{eff}$=1.5 eV in a Dudarev implementation\cite{Dudarev:1998aa} to treat the localized $d$ states in Co throughout the calculations except for the calculations which require the variation of the $U_{\mathrm{eff}}$ values. The result with $U_\mathrm{eff}$=1.5 eV for P-SCO give the FM moment of 2.6 $\mu_B$/f.u. and the lattice constant $a_0 = 3.840$ \AA\ in good agreement with previous theoretical\cite{Lee:2011aa} and experimental\cite{Long:2011aa} reports.

The energy barrier for structural transition from \textit{Pnma} to \textit{I2mb} space group have been computed using the nudged elastic band (NEB) method\cite{Mills:1995} as implemented in VASP.\cite{Henkelman:2000,Henkelman:2000a} Eight intermediate images are used to calculations.
The macroscopic polarization was calculated using Berry phase method.\cite{King-Smith:1993,Resta:1994}

% \section{Results}
\section{Perovskite Structure S\lowercase{r}C\lowercase{o}O$_{3-\delta}$}
\label{sec:p-sco}

\subsection{Electronic structure of stoichiometric SrCoO$_{3}$}
\label{sub:electronic structure of P-SCO}

\begin{figure}
	\centering
	\includegraphics[width=0.9\linewidth]{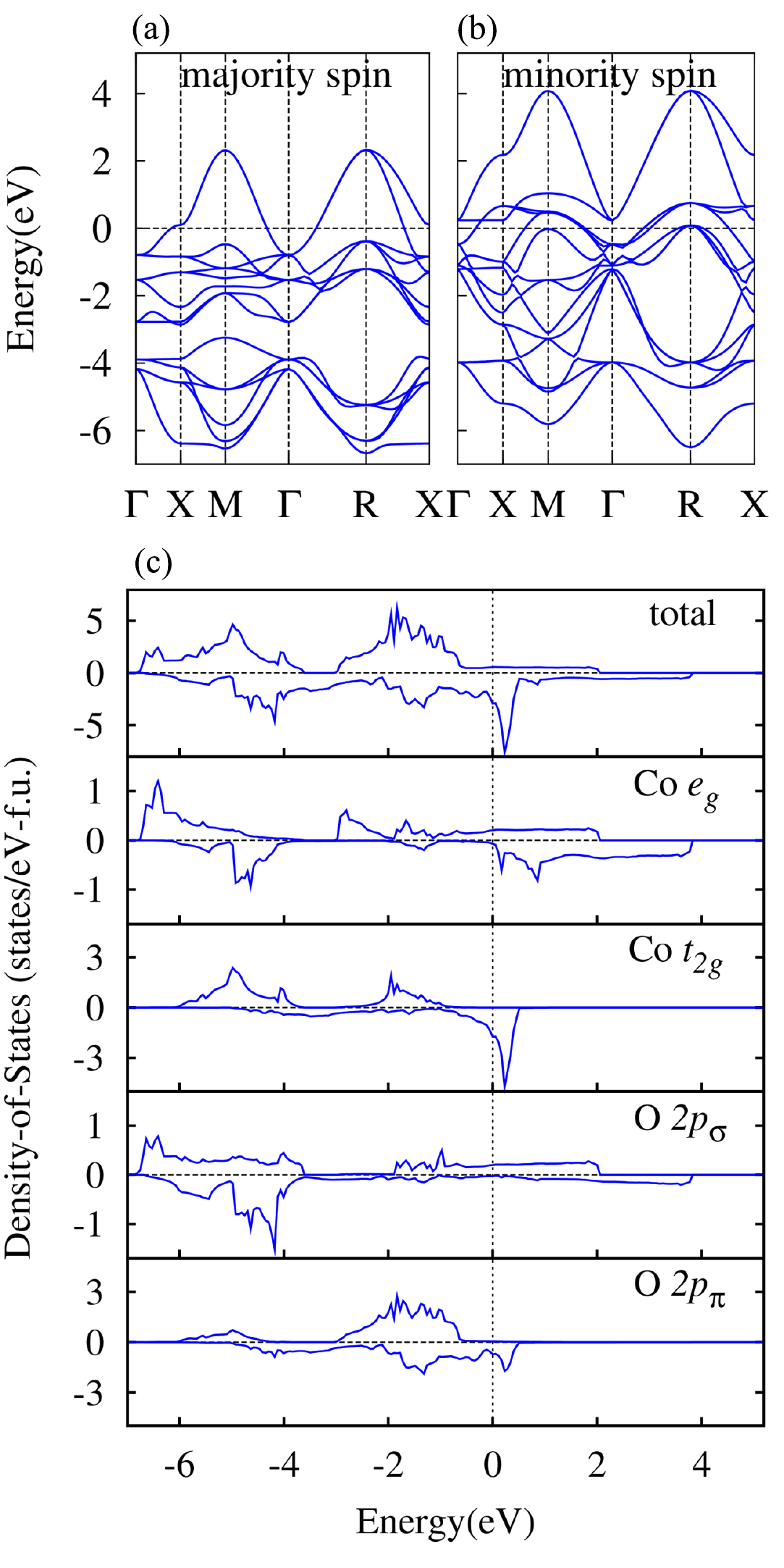}
	\caption{Band structures of (a) spin-up (majority-spin) and (b) spin-down (minority-spin) channels, and (c) the total and projected density-of-states (DOS) of perovskite SrCoO$_{3}$. In each DOS panel of (c), the positive (negative) scale in the vertical axis represents the DOS of spin-up (spin-down) components, respectively.}
	\label{fig:p-sco-banddos}
\end{figure}

The overall features of electronic band structures for stoichiometric SrCoO$_{3}$ (P-SCO) are in general agreement with previous works,\cite{Jaya:1991aa,Zhuang:1998aa,Ali:2013aa,Choi:2013aa,Hoffmann:2015aa} as shown in Fig.~\ref{fig:p-sco-banddos}. To calculate the electronic band structures and projected density-of-states (pDOS), we adopt the same parameter of $U_{\mathrm{eff}}=1.5$ eV as used in Ref.~\onlinecite{Lee:2011aa} to ensure the best fit to the ground state properties of lattice constant and magnetic moment for the FM metallic perovskite SrCoO$_{3}$.\cite{Long:2011aa}  The dispersive bonding and antibonding bands near $-5$ eV below the Fermi level ($E_{\mathrm{F}}$) and just above $E_{\mathrm{F}}$, respectively, in both majority spin (spin-up) and minority-spin (spin-down) channels of Fig.~\ref{fig:p-sco-banddos}(a) and (b) are the signature of a strong $dp\sigma$-hybridization between Co $e_{\mathrm{g}}$ and O $p_{\sigma}$ orbitals. The separation of the $dp\sigma$ bonding-antibonding levels is about 8 eV. It is much larger than that of the Co $t_{\mathrm{2g}}$ and O$p\pi$ bonding-antibonding levels, which is close to 4 eV.

As illustrated in the pDOS plot of Fig.~\ref{fig:p-sco-banddos}(c), the ferromagnetic ground state of P-SCO has a large exchange split of $\sim 5$ eV for the $t_{\mathrm{2g}}$ states so that the spin-up channel of the Co $t_{\mathrm{2g}}$-O $p_{\pi}$-hybridized bands is fully occupied while the spin-down channel is partially filled. Thus, its spin configuration can be assigned to $t_{\mathrm{2g}}^{3\uparrow\ 1\downarrow}$. Despite the large exchange split in $t_{\mathrm{2g}}$, however, the spin-polarization of the $dp\sigma$-hybridized bands are not pronounced because the $dp\sigma$ bonding-antibonding separation exceeds the exchange energy scale, and indeed counting the spin-polarization for the Co $e_{\mathrm{g}}$ component can be quite tricky. The spin-down component of O $p_{\sigma}^{\downarrow}$ is almost fully occupied with some fraction of the hybridized Co $e_{\mathrm{g}}^{\downarrow}$ states near -4.5 eV below $E_{\mathrm{F}}$. On the contrary, the spin-up component of the Co $e_{\mathrm{g}}$-O $p_{\sigma}$ antibonding band is nearly empty, which can be attributed to the ligand hole formed by the strong Co 3$d$ and O 2$p$ hybridization, as suggested by the earlier atomic multiplet calculation.\cite{Potze:1995aa} Thus, although it seems not easy to assign a single spin-configuration to the ground state of P-SCO, its spin-configuration can be matched approximately to an intermediate spin (IS) state close to a mixture of $t_{\mathrm{2g}}^{3\uparrow\ 1\downarrow}e_{\mathrm{g}}^{1\uparrow}$ and $t_{\mathrm{2g}}^{3\uparrow\ 1\downarrow}e_{\mathrm{g}}^{2\uparrow}\underbar{L}$.

The significance of the mixed-states of Co $d$ electrons and O $p$ holes in P-SCO has been emphasized by the atomic multiplet calculations\cite{Potze:1995aa} as well as other Hartree-Fock calculations.\cite{Takahashi:1998aa,Zhuang:1998aa} The competition between the crystal field strength and Hund’s coupling was found to be a key parameter for the Co spin state and showed that the change of crystal field strength drives the P-SCO system from the IS state to the LS state, still emphasizing that the IS state of Co $t_{\mathrm{2g}}^4 e_{\mathrm{g}}^1$ is the most probable candidate for P-SCO.\cite{Zhuang:1998aa} It is interesting to note that the itinerant holes residing in the O $p$ ligands, arising from the strong $pd$-hybridization, are coupled to neighboring Co ions antiferromagnetically, mediating the ferromagnetic ordering of Co local moments.\cite{Potze:1995aa} This mechanism shares a common feature with the Zener's mechanism\cite{Zener:1951aa} where an effective exchange interaction is generated by the $sd$-hybridization instead of the $pd$-hybridization. Basically, it is essential to have the energy gain produced by the negative polarization of the $p$ state, which is considered as a relaxation of the non-magnetic elements and eventually stabilizes the ferromagnetic ordering of Co spins.\cite{Kanamori:2001aa}

\subsection{$U$-dependent ground state of SrCoO$_{3-\delta}$}
\label{sub:u-dependent ground state}

While the mechanism behind the FM metallic ground state of P-SCO is subtle, the stability of this FM ground state has been shown to be fragile against the oxygen vacancy or lattice strain. There is experimental evidence that the change of oxygen content in SrCoO$_{3-\delta}$ results in a wide variety of electronic and magnetic properties from FM metal to AFM insulator as a pathway of topotactic transformation between P-SCO ($\delta$=0) and BM-SCO ($\delta$=0.5)\cite{Jeen:2013aa} and further the formation of oxygen vacancy mediated by the strain control can induce such FM-to-AFM phase transitions.\cite{Jeen:2013ab,Callori:2015aa,Petrie:2016aa,Hu:2017aa} There is also a first-principle study demonstrating that the formation of oxygen vacancy is favored by the volume expansion of the lattice in SrCoO$_{3-\delta}$.\cite{Cazorla:2017aa} All of the evidence point to that the oxygen vacancy $\delta$ is the single most important parameter modifying the Co valence state without cation doping which determines the electronic and magnetic properties of SrCoO$_{3-\delta}$.

However, on the other hand, recent first-principles calculations demonstrated that the epitaxial strain on P-SCO could also induce electronic and magnetic phase transitions from FM-metal to AFM-insulator,\cite{Lee:2011aa} which signifies that the lattice strain can be another important parameter controlling the magnetic state of Co ions in P-SCO. In fact, there are many theoretical indications that the Co spin state in P-SCO is in proximity to HS, LS, or even IS states with either FM or AFM ordering,\cite{Potze:1995aa} where its spin state depends on the competition between the crystal field strength and Hund’s coupling \cite{Zhuang:1998aa} and the presence of oxygen vacancy.\cite{Hoffmann:2015aa} Similar changes of the Co spin state in LaCoO$_{3}$ have been reported for the temperature\cite{Podlesnyak:2006aa,Haverkort:2006aa,Maris:2003aa,Klie:2007aa} as well as the lattice strain.\cite{Kwon:2014aa}

The ground-state electronic and magnetic structures of stoichiometric SrCoO$_{3}$ has recently been reexamined extensively based on density functional theory calculations with GGA+$U$ and hybrid functional by Rivero and Cazorla.\cite{Rivero:2016aa} They proposed a tetragonal phase with possible Jahn-Teller (JT) distortions as a ground state structure for the stoichiometric SrCoO$_{3}$. Despite that a larger value of $U$= 6 eV was used for the calculations, which may be attributed to the discrepancy between their work and the previous works,\cite{Lee:2011aa,Long:2011aa} it is interesting to observe that the on-site Coulomb interaction of $U$ has a strong effect on the structural, electronic, and magnetic properties of P-SCO.

\begin{figure}
	\includegraphics[width=0.9\linewidth]{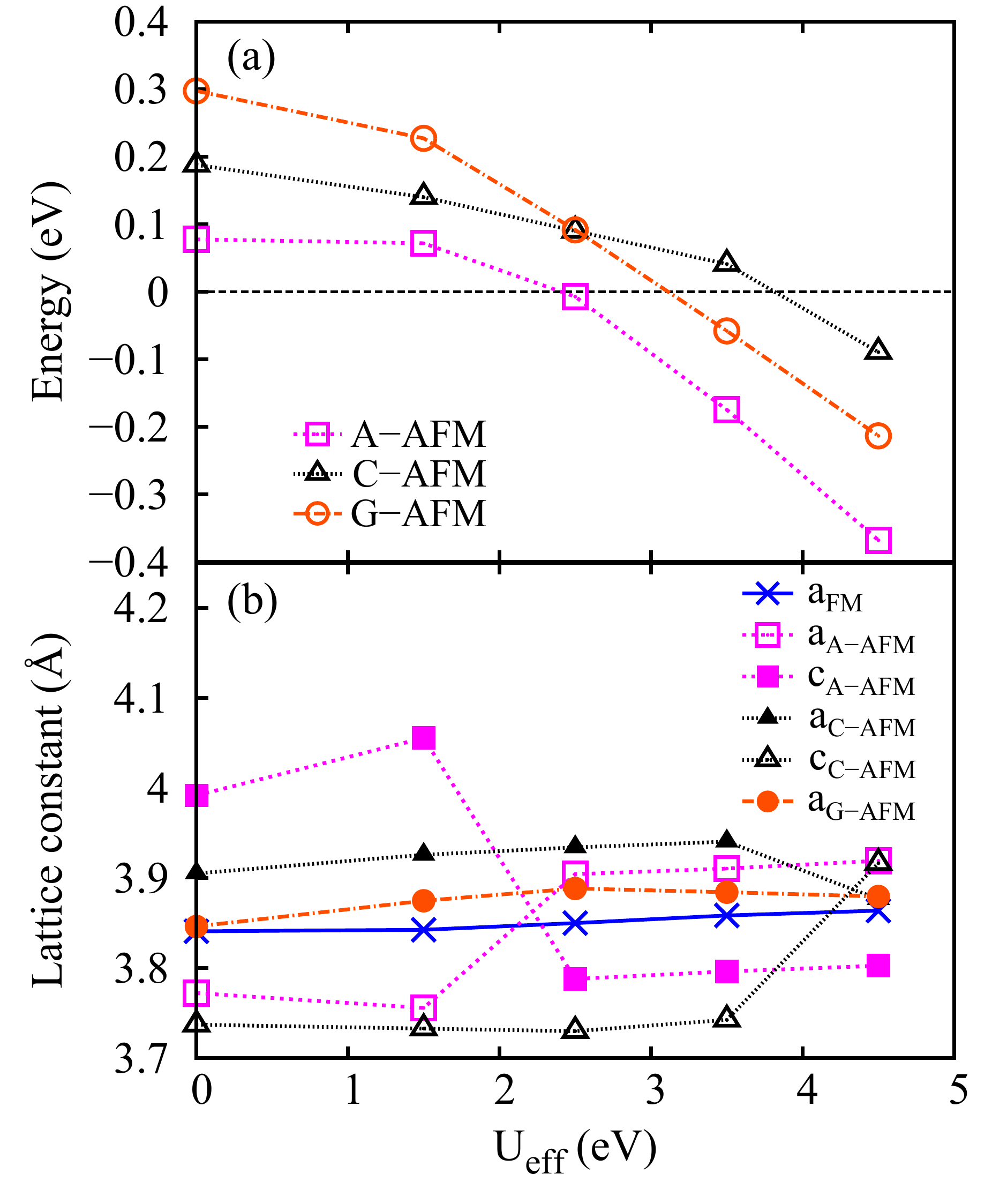}
	\caption{(Color online) $U$-dependent (a) total energies and (b) effective lattice constants for the spin configurations of stoichiometric SrCoO$_{3}$ with ferromagnetic (FM; cross (blue) mark), A-type (A-AFM; square (pink) mark), C-type (C-AFM; triangle (black) mark), and G-type (G-AFM; circle (orange) mark) antiferromagnetic orderings. The energy of each spin configuration in (a) is provided relative to that of the FM configuration. Solid symbols in (b) represent for the lattice constants along the Co-Co bonds with antiferromagnetic spin orderings, while open symbols for the Co-Co bonds with ferromagnetic spin orderings.}
	\label{fig:p-sco-u-dep}
\end{figure}

To investigate the effect of $U$ on the lattice structure and magnetic ordering, we performed GGA+$U$ calculations with varying the $U$ values and examined the equilibrium lattice structures and corresponding total energies.
Figure~\ref{fig:p-sco-u-dep}(a) illustrates the calculated total energies for several lowest-energy spin configurations of stoichiometric SrCoO$_{3}$ with ferromagnetic (FM), A-type (A-AFM), C-type (C-AFM), and G-type (G-AFM) antiferromagnetic orderings. At each value of $U$, the total energy for a given spin configuration is calculated after the full relaxation of \textit{equilibrium} lattice parameters as well as internal coordinates of atoms within the $\sqrt{2}\times\sqrt{2}\times 2$ cell of the SrCoO$_{3}$ formula unit, which can accommodate all the spin configurations under consideration.

The FM state of P-SCO, as shown in Fig.~\ref{fig:p-sco-u-dep}(a), is found to be the ground state up to the value of $U_{\mathrm{eff}} < 2.5$ eV, which is consistent with the results of previous calculations.\cite{Lee:2011aa} For the values of $U_{\mathrm{eff}} > 2.5$ eV, all the AFM ordered states becomes stable relative to the FM state. Among the AFM states, however, the most stable configuration is the A-AFM structure, where the Co spins order ferromagnetically within the $ab$-plane but antiferromagnetically along the $c$-axis, i.e., layer-by-layer. The lowering of crystal symmetry and the presence of JT  distortions in this A-AFM phase are well compared to the large-$U$ calculations by Rivero and Cazorla,\cite{Rivero:2016aa} demonstrating that the on-site Coulomb interactions play a crucial role in determining the electronic and magnetic structures of P-SCO.

Near the transition point of $U_{\mathrm{eff}} = 2.5$ eV, the total energies for all the different spin configurations are merging, and the FM ground state of P-SCO comes to close proximity to other AFM states. This behavior seems to be consistent with the observation that the P-SCO can undergo electronic and magnetic phase transitions with an extra epitaxial strain for the small $U$ values\cite{Lee:2011aa} and the proposed tetragonal phase of P-SCO with possible Jahn-Teller (JT) distortions for the larger value of $U$.\cite{Rivero:2016aa}

For the range of $U > 2.5$ eV, the lowest energy turns out to be the A-AFM state. From the $U$-dependent total energies for various spin configurations in Fig.~\ref{fig:p-sco-u-dep}, it is obvious that the increase of $U$ boosts the stability of AFM spin ordering. However, on the other hand, the G-AFM ordering is less favorable to the A-AFM ordering for all the range of $U$ considered. The band structure features of the G-AFM configuration of P-SCO are quite similar to that of BM-SCO, i.e., the G-AFM band structure of SrCoO$_{2.5}$, which is discussed in detail in Fig.~\ref{fig:bm-banddos1} of Sec.~\ref{sub:structure-electronic-BM-SCO}. A notable characteristic is the \emph{strong} suppression of the $dp\sigma$-hybridization between Co $e_{\mathrm{g}}$ and O $p_{\sigma}$ orbitals. As results, the partially-filled $dp\sigma$ anti-bonding bands in Fig.~\ref{fig:p-sco-banddos}(a), which is essential to the stability of the FM state, are no longer present near the Fermi level and, further, contribute to the increase of the Co localized moment. Consequently, when all the neighboring spins are antiferromagnetically ordered, a finite bandgap opens up only for the G-AFM configuration of P-SCO for $U > 2.5$ eV.

\subsection{Effective Co-Co distance and magnetic orderings}
\label{sub:effective-co-co-distance}

Another interesting aspect emerges in the relation between the spin-ordering and the Co-O bond length as a function of $U$. The lattice constants in Fig.~\ref{fig:p-sco-u-dep}(b) for P-SCO represents the distance between neighboring Co atoms in each spin configuration, which is proportional to the Co-O bond length. As for the FM and G-AFM cases, all the bondings are equivalent and give a single parameter. On the other hand, as for the A-AFM and C-AFM configurations, the bond length shows a large variation depending on the spin configuration. Overall, the bond lengths between AFM spins are significantly longer than FM. At $U=0$ eV, for example, $c_{\mathrm{A-AFM}}= 4$ {\AA} for the Co-Co distance between AFM-ordered layers, while $a_{\mathrm{A-AFM}}= 3.78$ {\AA} for the Co-Co distance within the FM layer. Since the larger Co-Co distance gives rise to a smaller $dp\sigma$ hybridization, its contribution to the FM order will be reduced significantly, thereby leading to the AFM order.

Incidentally, in Fig.~\ref{fig:p-sco-u-dep}(b), one can observe interesting crossovers happening between AFM- and FM-bond lengths at $U=2.5$ eV for A-AFM and $U=4.5$ eV for C-AFM, respectively. At a glance, it appears to be contradicting to the general trend of the Co-Co distance and the neighboring spin order, namely, the longer bond for AFM and the shorter bond for FM. But, after examining the electronic structures across the transition from $c_{\mathrm{A-AFM}}>a_{\mathrm{A-AFM}}$ to $c_{\mathrm{A-AFM}}<a_{\mathrm{A-AFM}}$, we find that this crossover behavior is triggered by the JT distortion within the $t_{\mathrm{2g}}$ manifold in the minority spin channel.

For the range of $c_{\mathrm{A-AFM}}>a_{\mathrm{A-AFM}}$, the Co $d_{yz,zx}$ orbital states are degenerate and lower in energy than that of Co $d_{xy}$ so that they contribute to the partially occupied $t_{\mathrm{2g}}$ band at $E_{\mathrm{F}}$, similarly to the Co $t_{\mathrm{2g}}$ panel of P-SCO in Fig.~\ref{fig:p-sco-banddos}(c). It means that these  Co $d_{yz,zx}$ orbitals form a conducting channel in the FM layers of the A-AFM configuration. Therefore, the shorter Co-Co distance within the $ab$-plane favors the FM ordering energetically. In the case of $c_{\mathrm{A-AFM}}<a_{\mathrm{A-AFM}}$, however, the JT distortion triggers the local distortion of CoO$_{6}$ octahedron for $c_{\mathrm{A-AFM}}<a_{\mathrm{A-AFM}}$ reverses the crystal field splitting, brings down the single Co $d_{xy}$ level below $E_{\mathrm{F}}$ and generates a energy gap between the fully occupied $d_{xy}$ and empty $d_{yz,zx}$ bands.

Although the intervention of the JT distortion for the larger value of $U$ makes the electronic and magnetic structure of P-SCO even more complicated, we can conclude that the AFM ordering is a result of the competition between the on-site Coulomb interaction $U$ and the $dp\sigma$ hybridization. Further, the electronic band structure depends on the ordering of neighboring Co spins, which is strongly affected by the Co-O bond length. The antiferromagnetic ordering between neighboring Co spins is crucial to the bandgap formation, but the energy gain by the AFM order is not strong enough to destroy the stability of FM configurations sustained by the presence of strong $dp\sigma$ hybridization between Co $e_{\mathrm{g}}$ and O $p_{\sigma}$ orbitals as evidenced in Fig.~\ref{fig:p-sco-u-dep}(a).

\subsection{Oxygen vacancy and effective Co-Co distance}
\label{sub:oxygen-vac-co-co-distance}

\begin{figure}
	\includegraphics[width=0.9\linewidth]{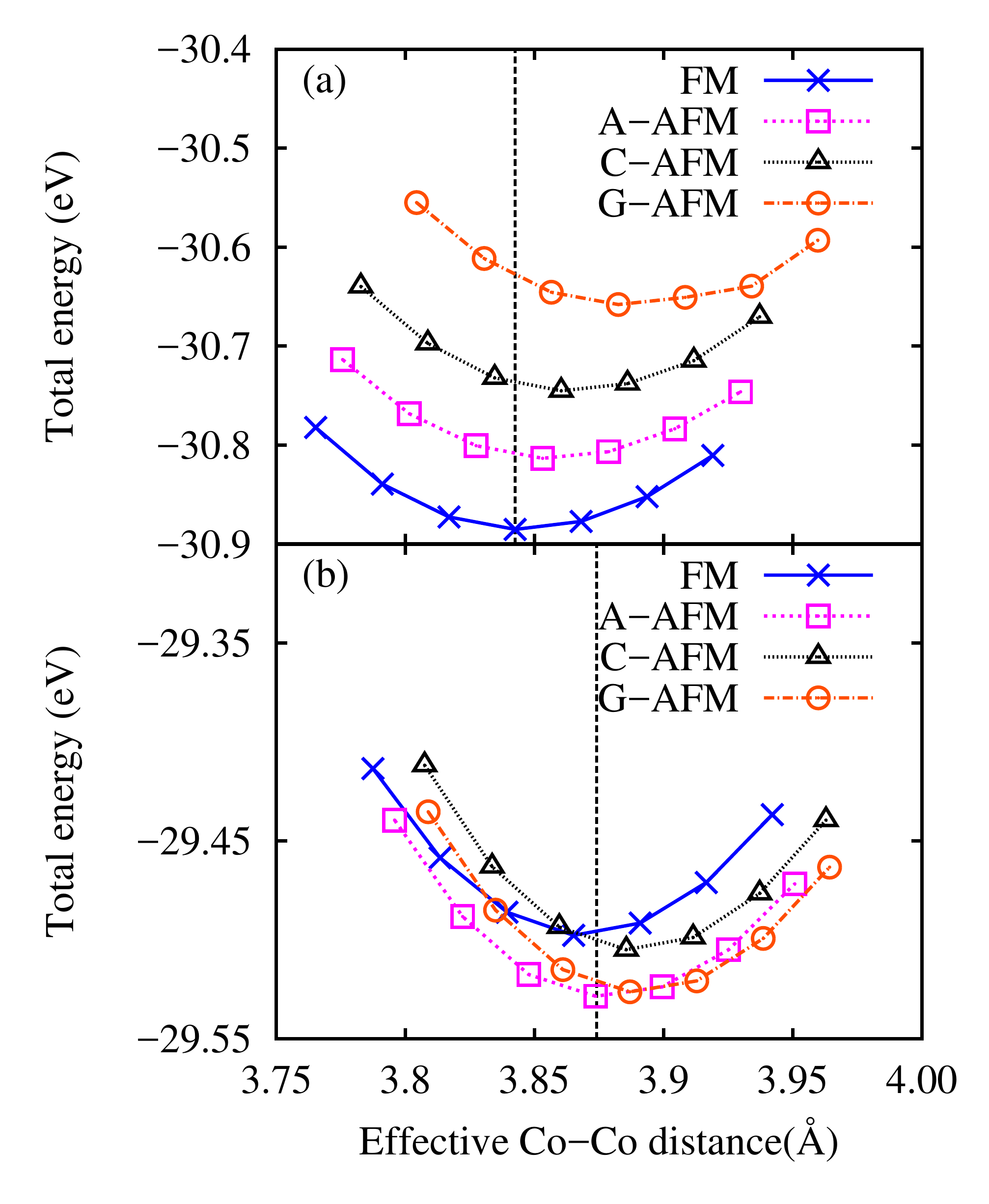}
	\caption{(Color online) Total energy vs. effective Co-Co distance for different magnetic configurations of (a) stoichiometric SrCoO$_{3}$ (P-SCO) and (b) SrCoO$_{2.75}$ (SCO-V0). Spin configuration are marked by crosses for FM, squares for A-AFM, triangles for C-AFM, and circles for G-AFM, respectively.}
	\label{fig:co-co-distance}
\end{figure}

It is well known that the volume of transition metal oxides increases when oxygen vacancies are introduced.\cite{Hoffmann:2015aa,Adler:2004,Chen:2005} The SrCoO$_{3-\delta}$ also exhibits a volume increase, which corresponds to an increase of the average Co-Co distance by about 2\% from P-SCO to BM-SCO. This difference is indeed larger than the difference of effective lattice constants between FM and G-AFM of P-SCO found in Fig.~\ref{fig:p-sco-u-dep}(b). To gain an insight into the relation between lattice constants and their magnetic ordering, we performed DFT calculations to determine total energies, as a function of the effective Co-Co distance, for the stoichiometric SrCoO$_{3}$ (P-SCO) and SrCoO$_{2.75}$ (SCO-V0) in several low-energy spin configurations, as illustrated in Fig.~\ref{fig:co-co-distance}. As for the SCO-V0 structure, we introduce one oxygen vacancy within the $\sqrt{2}\times\sqrt{2}\times2$ supercell of the SrCoO$_{3}$ formula unit, which comprise SrCoO$_{2.75}$. It should be noted that this SCO-V0 structure does not have an oxygen-vacancy ordering similar to that of BM-SCO. For all the calculations, all the lattice constant as well as the internal position of atoms are optimized without any symmetry constraint except the given spin configuration. The effective Co-Co distance is drawn from the averaged volume per Co atom for each configuration.

Figure~\ref{fig:co-co-distance} demonstrates that all the structures with AFM-ordering have larger lattice volumes, i.e., the effective Co-Co distance, than those of the FM-ordered structures in both P-SCO and SCO-V0, respectively. In fact, the Co-Co distances increase as the number of AFM neighbors grows from FM to A-AFM to C-AFM to G-AFM for both P-SCO and SCO-V0 cases. The spread of the equilibrium distances for different spin configurations shrinks for SCO-V0, which may be associated with the local lattice distortions induced by oxygen vacancies. Nonetheless, these behaviors are consistent with the general trend of the Co-Co distance and the neighboring spin order discussed in Fig.~\ref{fig:p-sco-u-dep}, namely, the long bond length for the AFM Co-Co neighbors and the shorter for FM. This leads to an interesting conjecture that the increase of lattice volume, i.e., the effective Co-Co distance, determines the magnetic ground state by means of controlling the effective $dp\sigma$ hybridization between Co $e_{\mathrm{g}}$ and O $p\sigma$ orbitals.

\section{Brownmillerite Structure S\lowercase{r}C\lowercase{o}O$_{2.5}$}
\label{sec:brownmillerite-sco}

\subsection{Structural and electronic properties of brownmillerite SrCoO$_{2.5}$}
\label{sub:structure-electronic-BM-SCO}

The brownmillerite SrCoO$_{2.5}$ (BM-SCO) crystal has an orthorhombic structure and consists of alternating layers of CoO$_{6}$ octahedra and CoO$_{4}$ tetrahedra.\cite{Toquin:2006aa,Sullivan:2011aa} The unit cell of BM-SCO is characterized by the Co(1) at the octahedral site and the Co(2) at the tetrahedral site. Here we follow the notations used in work by Mu\~{n}os et al.\cite{Munoz:2008aa} The Co(1) atom inherits the original octahedral environment of Co from P-SCO, while the Co(2) atom has distorted tetrahedral coordination of oxygen atoms. The layer of CoO$_{4}$ tetrahedra can be viewed by the ordered line of oxygen vacancies along the $[1\bar{1}0]$ direction in every second $(00l)$ layer of stoichiometric SrCoO$_{3}$ (P-SCO).

\begin{table}
	\caption{Calculated lattice structures for SrCoO$_{3}$ and SrCoO$_{2.5}$. The unit is in {\AA}. O$_\mathrm{oct}$ is the oxygen atom located within the octahedral layer and O$_\mathrm{tet}$ is the oxygen atom within the tetrahedral layer. O$_\mathrm{int}$ represents the oxygen atom located between the octahedral and tetrahedral layers.}
	\label{tab:strdata}

	%\centering
	\begin{ruledtabular}
		\begin{tabular}{l l l l}
			%\hline\hline
			\multicolumn{2}{l}{Compound}                  & Calculation            & Experiment                                             \\
			\hline
			\multirow{2}{*}{SrCoO$_3$}                    & Lattice constant       & 3.843        & 3.829\footnote{Ref.\cite{Long:2011aa}}  \\
			                                              & Co-O                   & 1.921        & 1.915                                   \\
			\hline
			\multirow{7}{*}{SrCoO$_{2.5}$(\textit{I2mb})} & Lattice constant $a$   & 5.532        & 5.458\footnote{Ref.\cite{Munoz:2008aa}} \\
			                                              & $b$                    &  15.85       & 15.64                                  \\
			                                              & $c$                    &  5.639       &  5.564                                   \\
			                                              & Co(1)-O$_\mathrm{oct}$ & 2.101, 1.848 & 1.963, 1.938                            \\
			                                              & Co(2)-O$_\mathrm{tet}$ & 1.922, 1.913 & 2.241, 1.816                            \\
			                                              & Co(1)-O$_\mathrm{int}$ & 2.283        & 2.214                                   \\
			                                              & Co(2)-O$_\mathrm{int}$ & 1.824        & 1.801                                   \\ \hline
			\multirow{7}{*}{SrCoO$_{2.5}$(\textit{Pnma})} & Lattice constant $a$   & 5.523        & 5.458\footnotemark[2]                   \\
			                                              & $b$                    & 15.83        & 15.64                                   \\
			                                              & $c$                    & 5.631        & 5.564                                   \\
			                                              & Co(1)-O$_\mathrm{oct}$ & 2.098, 1.853 & 1.990, 1.912                            \\
			                                              & Co(2)-O$_\mathrm{tet}$ & 1.922, 1.912 & 2.200, 1.828                            \\
			                                              & Co(1)-O$_\mathrm{int}$ & 2.262        & 2.216                                   \\
			                                              & Co(2)-O$_\mathrm{int}$ & 1.827        & 1.797                                   \\
			%\hline\hline
		\end{tabular}
	\end{ruledtabular}
	%$^a$Ref.~\onlinecite{Long:2011aa}, $^b$Ref.~\onlinecite{Munoz:2008aa}\\
\end{table}

Both the experimental and calculated values of the lattice constants and the bond lengths of P-SCO and BM-SCO are listed in Table~\ref{tab:strdata}. For the brownmillerite structure of SrCoO$_{2.5}$, we consider two possible space groups of \textit{I2mb} and \textit{Pnma}, the structures of which depend on the stacking configuration of vacancy orderings in each tetrahedral layer.	To determine the ground state structure of SrCoO$_{2.5}$, the experimental structures of both \textit{I2mb} and \textit{Pnma} were used for further relaxations of the lattice constants and internal coordinates as well. The lattice constants obtained after the relaxation are $a=5.532$ {\AA}, $b=15.85$ {\AA}, $c=5.639$ {\AA} for \textit{I2mb} and $a=5.523$ {\AA} , $b=15.83$ {\AA} , $c=5.631$ {\AA} for \textit{Pnma}. The total energy of \textit{I2mb} is found to be lower than \textit{Pnma} by a few meV per formula unit regardless of the values of $U_\mathrm{eff}$ used in the calculations. With the choice of  $U_{\mathrm{eff}}=1.5$ eV, the lattice constant and Co-O distance for P-SCO is in excellent agreement with experiment.\cite{Long:2011aa} However, the calculated lattice constants for BM-SCO turn out to be slightly larger (more than 1 {\%}) than experimental results.\cite{Munoz:2008aa} The main discrepancy between calculations and experiment lies mostly in the internal positions of oxygens.

This mismatch between our calculation and experiment is rather significant. During the relaxation of the internal coordinates of atoms in BM-SCO, we observe that a large difference between the short and long bond lengths of Co(1)-O$_\mathrm{oct}$ arises from the JT distortion of the Co(1) octahedron within the octahedral layers, which was not reported in the experiment.\cite{Munoz:2008aa} Indeed, this JT distortion in the octahedral layer is compatible with the observed \textit{I2mb} space group symmetry and needs to be confirmed by experiment in the future. More discussion will be given in Sec.~\ref{sub:ferroelectricity in BM-SCO} concerning the tetrahedral chain ordering and ferroelectric in the tetrahedral layers of Co(2).

\begin{figure*}
	\centering
	\includegraphics[width=0.9\linewidth]{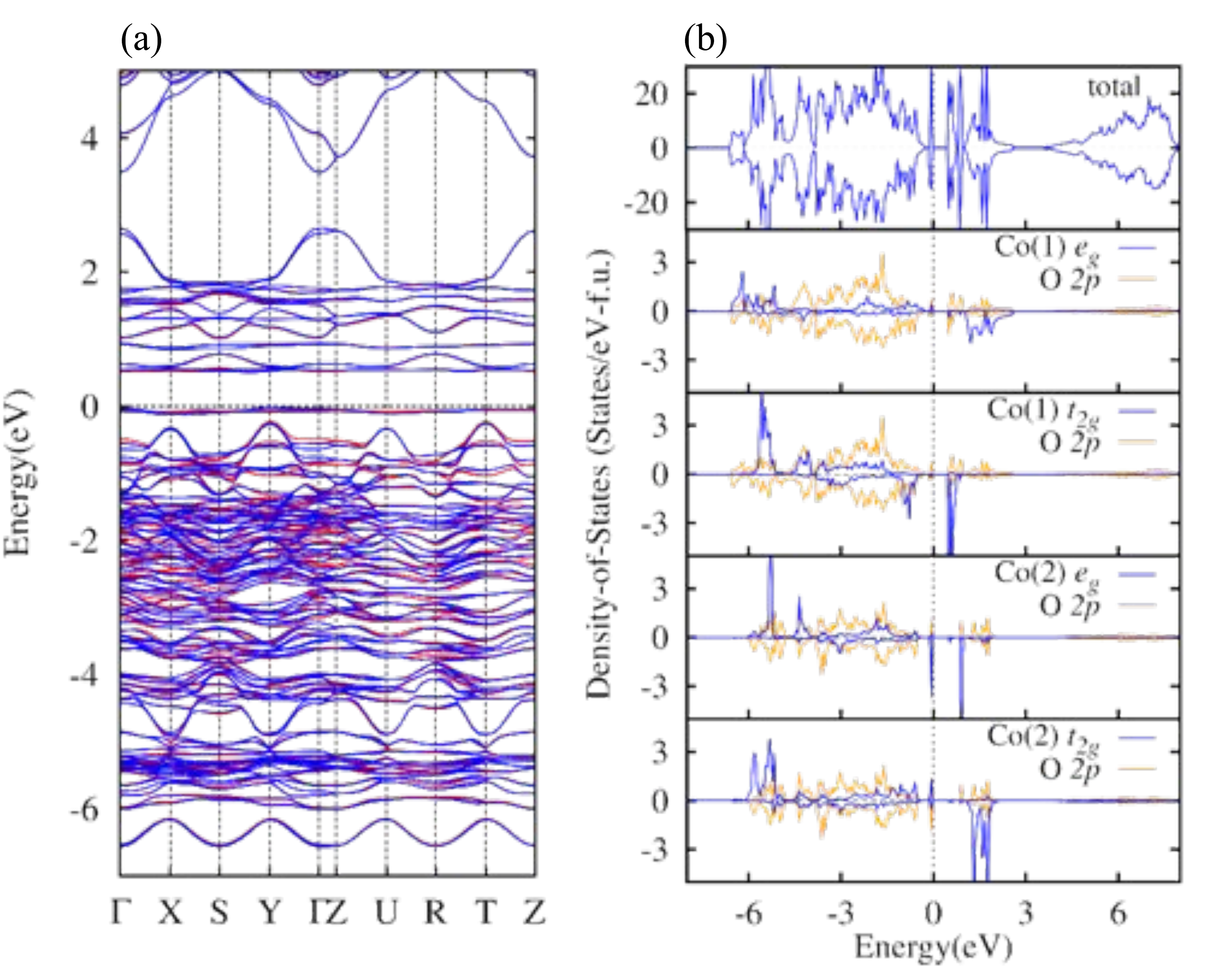}
	\caption{(Color online) (a) Spin-polarized band structure and (b) projected density-of-states (DOS) for the G-AFM ground state of brownmillerite structure SrCoO$_{2.5}$ with U$_\mathrm{eff}$=1.5 eV. The spin-up and spin-down bands in (a), illustrated by red and blue lines, are nearly degenerate. The non-degenerate states near the Fermi level are related to the ferrolectric ordering breaking the inversion symmetry. (See the text for details.) In each DOS panel of (b), the positive (negative) scale in the vertical axis represents the DOS of spin-up (spin-down) components, respectively. The pDOS of Co ions and O ions are shown with dark (blue) and light (yellow) solid lines, respectively.}
	\label{fig:bm-banddos1}
\end{figure*}

Figure~\ref{fig:bm-banddos1} shows the calculated band structure and density-of-states for the G-AFM ground state of brownmillerite SrCoO$_{2.5}$ (BM-SCO). The calculated G-AFM magnetic ordering in BM-SCO is consistent with experiments.\cite{Sullivan:2011aa} The total energy of G-AFM also remains robust against other magnetic configurations even for the large values of $U_{\mathrm{eff}}$ up to 3.5 eV. The details of effective exchange interactions will be discussed in Sec.~\ref{sub:effective-exchange-BM-SCO}. The bandgap is about 0.5 eV, which is also consistent with the experimental direct band gap of $0.35 \sim 0.45$ eV.\cite{Choi:2013aa}

The features of the BM-SCO band structure are in general consistent with previous calculations.\cite{Munoz:2008aa,Pardo:2008aa,Mitra:2013aa,Choi:2013aa} In comparison with the band structure of P-SCO in Fig.~\ref{fig:p-sco-banddos}, one of the most prominent changes is the \emph{strong} suppression of the $dp\sigma$-hybridization between Co $e_{\mathrm{g}}$ and O $p_{\sigma}$ orbitals for both Co(1) and Co(2). The $dp\sigma$ anti-bonding bands of P-SCO in Fig.~\ref{fig:bm-banddos1}(a) are pushed up at $\sim 2$ eV above $E_{\mathrm{F}}$ for BM-SCO, as shown in Fig.~\ref{fig:bm-banddos1}(a), and their band-width is significantly reduced to $\sim 1$ eV, which is much smaller than that ($\sim 3$ eV) of P-SCO. This change of the $dp\sigma$-hybridization explains the stability of the AFM ordering even for the octahedrally coordinated Co(1) atoms regardless of the strength of on-site Coulomb interactions.

The detailed features of Co $d$-O $p$ hybridizations at the octahedral site of Co(1) and the tetrahedral site of Co(2) are illustrated in the pDOS plots of Fig.~\ref{fig:bm-banddos1}(b). It is noted that the pDOS panels of Fig.~\ref{fig:bm-banddos1}(b) display only either spin-up or spin-down atom among the same type of Co(1) or Co(2) atoms, respectively. The suppression of the $dp\sigma$-hybridization for Co(1) is well represented in the panel showing the pDOS of Co(1) $e_{\mathrm{g}}$-O 2$p$. In contrast to the strong mixture of Co $d$ electrons and O $p$ holes in P-SCO, all the spin-up $e_{\mathrm{g}}$ electrons are fully occupied, leaving almost no O $p$ hole except a minimal trace of orbital overlaps. Combining the pDOS results for the Co(1) $e_{\mathrm{g}}$ and $t_{\mathrm{2g}}$ components, we can identify that Co(1) has a valence state of Co$^{3+}$ $d^{6}$ in the high-spin (HS) state of $S=2$ with the configuration of $e_{\mathrm{g}}^{2\uparrow}t_{\mathrm{2g}}^{3\uparrow 1\downarrow}$. In case of Co(2), the $dp\sigma$-hybridization is even more suppressed. Since the order of $e_{\mathrm{g}}$ and $t_{\mathrm{2g}}$ levels is reversed in the tetrahedral coordination, as shown in the Co(1) $d$-O 2$p$ panels of Fig.~\ref{fig:bm-banddos1}(b), the valence state of Co(2) corresponds to Co$^{3+}$ $d^{6}$ in the high-spin (HS) state with $e_{\mathrm{g}}^{2\uparrow 1\downarrow}t_{\mathrm{2g}}^{3\uparrow}$.

Another unusual characteristic in the band structure of BM-SCO is the appearance of a localized Co(2) $e_{\mathrm{g}}$ state pinned just below $E_{\mathrm{F}}$. This localized state arises due to the absence of $dp\sigma$-hybridization in the tetrahedral environment of Co(2). Since this flat band forms a bandgap with the unoccupied states of Co(1) $t_{\mathrm{2g}}$ and Co(2) $e_{\mathrm{g}}$ states at a higher energy level, the interband transition across the bandgap can be significant. In fact, a similar feature has been observed and assigned to the local electronic structure of Co(2) by Choi \textit{et al.}\cite{Choi:2013aa}

\subsection{Tetrahedral chain ordering and ferroelectricity}
\label{sub:ferroelectricity in BM-SCO}

\begin{figure}
	\includegraphics[width=0.9\linewidth]{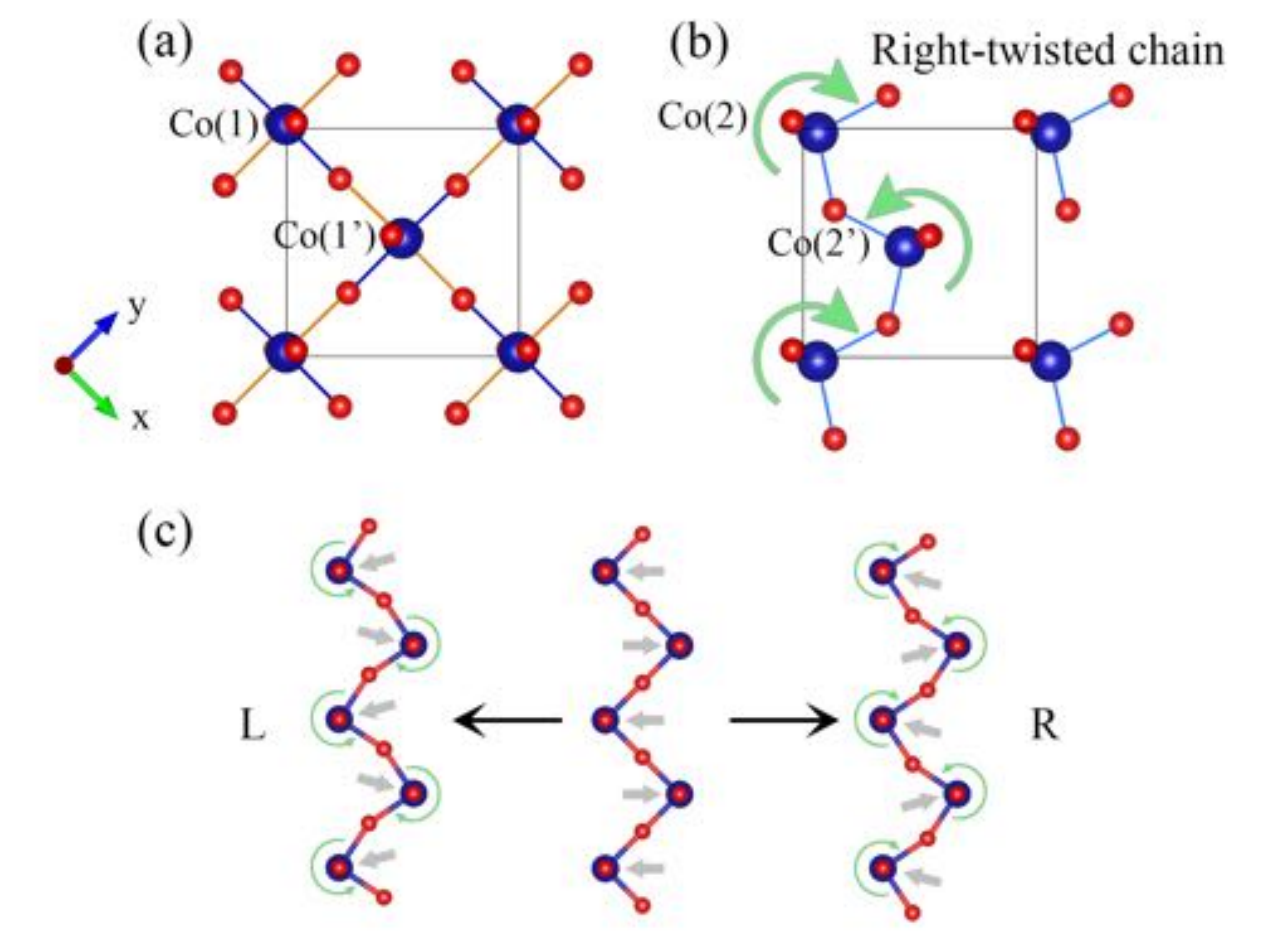}
	\caption{(Color online) Fully relaxed in-plane structures for the (a) octahedral (Co(1)) and (b) tetrahedral (Co(2)) sites. The lines with the same color represents the equivalent bonds in each layer. The circling arrows in (b) represent the rotation distortion of each tetrahedron of the R-twisted configuration. A schematic drawing of tetrahedral chains in (c) illustrates that the electric dipole moments of tetrahedrons cancel each other in the chain without any rotation while the R(L)-twisted chain develops a net moment parallel (antiparallel) to the $[\bar{1}10]$ direction, respectively.}
	\label{fig:bm-lattice}
\end{figure}

Figure~\ref{fig:bm-lattice}(a) and (b) illustrates the fully relaxed in-plane positions of Co and O atoms within the (a) octahedral layers and (b) tetrahedral layers. We introduce a local coordinate of the $x$- and $y$-axes in Fig.~\ref{fig:bm-lattice}(a), rotated by 45$^\circ$ to fit it into the bonding direction of the Co(1) octahedron for the description of $d$-electron orbitals. From Table~\ref{tab:strdata}, it is obvious that the pronounced prolate elongation happens to the apical bond between Co(1)-O$_{\mathrm{int}}$, which may be due to the presence of oxygen vacancy in the neighboring tetrahedral layers. In the prolate distortion, the $d_{xy}$ energy level stays higher than those of $d_{yz,zx}$ orbitals. It means that one spin-down $t_{\mathrm{2g}}^{1\downarrow}$ electron in the HS state of Co$^{3+}$ $d^{6}$ ($e_{\mathrm{g}}^{2\uparrow}t_{\mathrm{2g}}^{3\uparrow 1\downarrow}$) will occupy the doubly degenerate $d_{yz,zx}$ states and form a half-filled metallic band. When the in-plane JT distortion breaks the symmetry between the Co(1)-O bonds within the octahedral layer, as shown in Fig.~\ref{fig:bm-lattice}(a), one of the $d_{yz,zx}$ orbitals has lower energy and the degeneracy is removed. Consequently, for example, the $d_{yz}$-band of Co(1) becomes fully occupied and form a large gap as shown in the Co(1) $t_{\mathrm{2g}}$ - O 2$p$ pDOS panel of Fig.~\ref{fig:bm-banddos1}(b). Here it is noted that this pattern of the JT distortions of Co(1) and Co(1') atoms repeats in the second octahedral layers with a mirror inversion with respect to the plane parallel to [$\bar{1}10$] because of the non-symmorphic operation of the \textit{I2mb} space group, but does not affect the formation of the bandgap for Co(1) and Co(1').

Apart from the G-AFM insulating nature of BM-SCO, the tetrahedral layers consisting of oxygen vacancies are of interest. There are several works\cite{Abakumov:2005ab,Casey:2006aa,DHondt:2008aa,Parsons:2009aa} suggesting the chain ordering of the tetrahedral layers in other brownmillerite structured materials. According to Parsons \textit{et al.},\cite{Parsons:2009aa} the space group symmetry of brownmillerite structures can have 5 different categories: \textit{I2mb}, \textit{Pnma}, \textit{Pcmb}, \textit{C2/c} and \textit{Imma} depending on the inter-layer or intra-layer chain ordering of CoO$_{4}$ tetrahedra. Mu\~{n}os \textit{et al.}\cite{Munoz:2008aa} has proposed the space groups \textit{I2mb} and \textit{Pnma} for the SrCoO$_{2.5}$ brownmillerite structure. However, a recent experimental work did not find any long-range order in the tetrahedral layers and suggested the space group of \textit{Imma} for the disordered arrangement of CoO$_{4}$ tetrahedra.\cite{Sullivan:2011aa}

Among the five possible configurations of the tetrahedral ordering suggested by Parsons \textit{et al.},\cite{Parsons:2009aa} only the tetrahedral ordering with the \textit{I2mb} space group has a macroscopic electric polarization. The rotation of the tetrahedra results in two mirror-related configurations of the tetrahedral chains, which are arbitrarily called ``left'' (L) and ``right'' (R), depending on the sequence of tetrahedron rotations, as illustrated in Fig.~\ref{fig:bm-lattice}(c). In the \textit{I2mb} configuration shown in Fig.~\ref{fig:bm-lattice}(b), all the chains are found to have the same R-twisted configuration, consisting of the right-handed rotation of Co(2) and left-handed rotation of Co(2'). Since each rotation distortion breaks an inversion symmetry, however, the chains with the rotated tetrahedra acquire an electric dipole moment. The ordering of such tetrahedra chains conforms to the \textit{I2mb} symmetry and give rise to a ferroelectric moment along the chain direction. The calculated electric polarization of BM-SCO with the \textit{I2mb} symmetry is about 6.38 $\mu$C/cm$^{2}$ along the chain direction. In addition, there appears a relatively small component of polarization, $\sim$0.65 $\mu$C/cm$^{2}$, along with the $z$-direction, i.e., perpendicular to the layers, which is not forbidden by the crystal symmetry.

In general, each tetrahedral chain in the brownmillerite lattice can be changed into either L or R configurations.  In fact, the L or R configurations of the chains between tetrahedral layers (inter-layer) and within the tetrahedral layers (intra-layer) have been suggested to be the origin of the complexity of the brownmillerite-type structures.\cite{DHondt:2008aa} For example, the \textit{I2mb} symmetry corresponds to the configuration in which all tetrahedral chains are R-rotated, while the \textit{Pnma} structure has an alternating sequence of the R- and L-rotated chains. In \textit{Pnma}, the polarization of one tetrahedral chain with the R-rotated tetrahedra is compensated by the opposite polarization of the neighboring tetrahedral chain with L-twist. Thus, if the handedness of rotation is disordered between the layers or within the layer, the ferroelectric polarization of BM-SCO is expected to diminish.

\begin{figure}
  \includegraphics[width=0.9\linewidth]{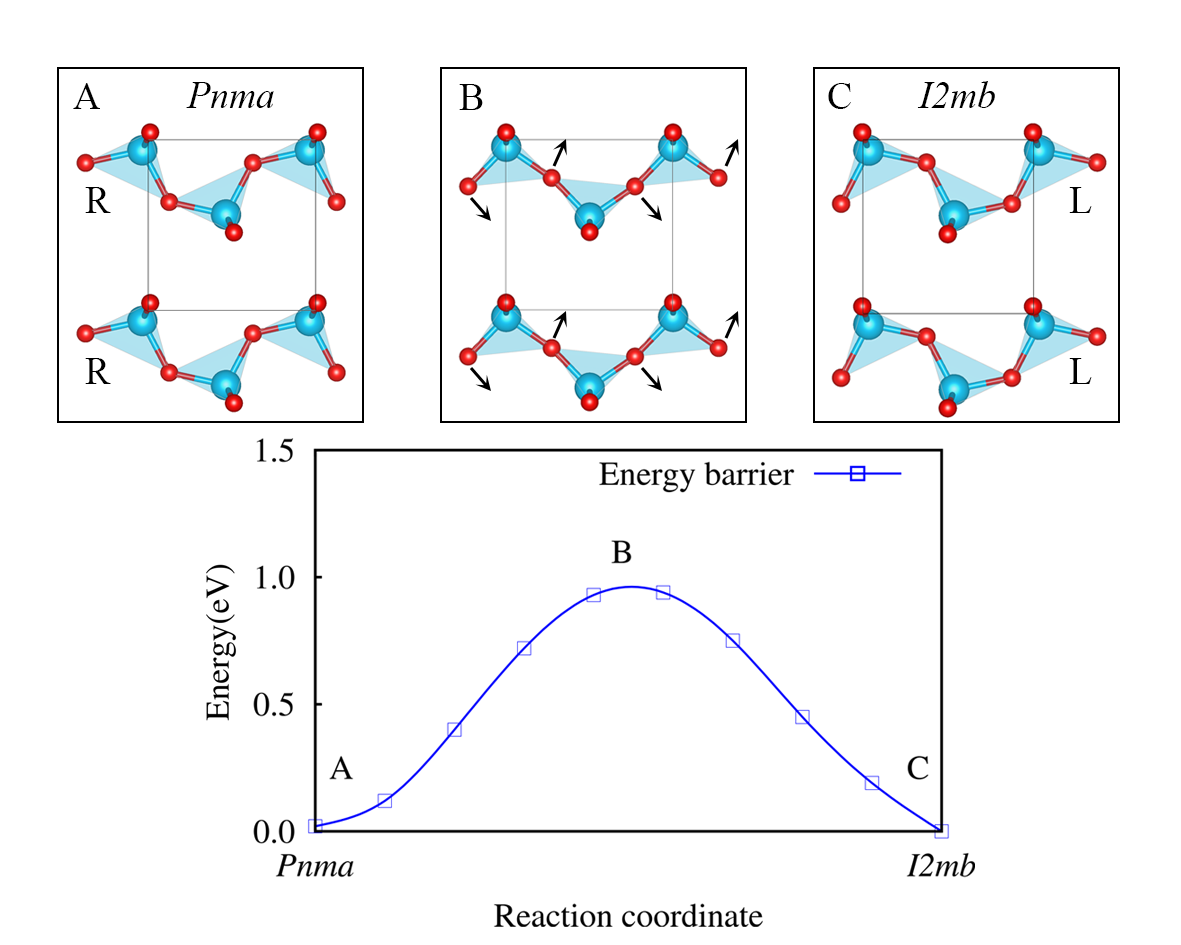}
  \caption{(Color online) Energy barrier across the structural transformation between \textit{Pnma} to \textit{I2mb} structures from the results of NEB calculations. The insets show a pathway connecting the R-twisted and L-twisted chains in one of the tetrahedral layers: (A) the R-twisted chain configuration of \textit{Pnma}, (B) the transient-state between \textit{Pnma} to \textit{I2mb}, and (C) the L-twisted chain configuration of \textit{I2mb}. The arrows in the inset indicate the direction of oxygen motion from \textit{Pnma} to \textit{I2mb}.}
  \label{fig:neb}
\end{figure}

The ferroelectricity, i.e., \textit{I2mb}, phase of BM-SCO is at the energy minimum relative to the other disordered configurations including the stacking disorder of \textit{Pnma}, but the energy difference between \textit{I2mb} and \textit{Pnma} is only 2.6 \textit{m}eV/f.u. Since the energy differences are so small, it is impossible to get the ordering by the thermodynamic annealing process. However, once the polarization of tetrahedral chains is aligned by a poling process, the stability of the \textit{I2mb} structure can be maintained by the activation barrier of the rotation of tetrahedral chains. To confirm this possibility, we calculated the activation barrier by using the nudged elastic band (NEB) method. Two endpoints of the configurations: A (\textit{Pnma}) and C (\textit{I2mb}) are shown in Fig.~\ref{fig:neb}. The insets of A, B and C show only one of the two independent tetrahedral layers, where A (\textit{Pnma}) has the L-rotation of tetrahedra and C (\textit{I2mb}) has the R-rotation. The inset B shows the direction of oxygen motion from A to C configurations. As results, the activation barrier for the transition between (\textit{Pnma}) and (\textit{I2mb}) is obtained to be $\sim 1$ eV, which is much larger than the room-temperature energy, suggesting a possible ferroelectric ordering for brownmillerite SrCoO$_{2.5}$.

\subsection{Effective exchange interactions in BM-SCO}
\label{sub:effective-exchange-BM-SCO}

\begin{figure*}
  \includegraphics[width=0.9\linewidth]{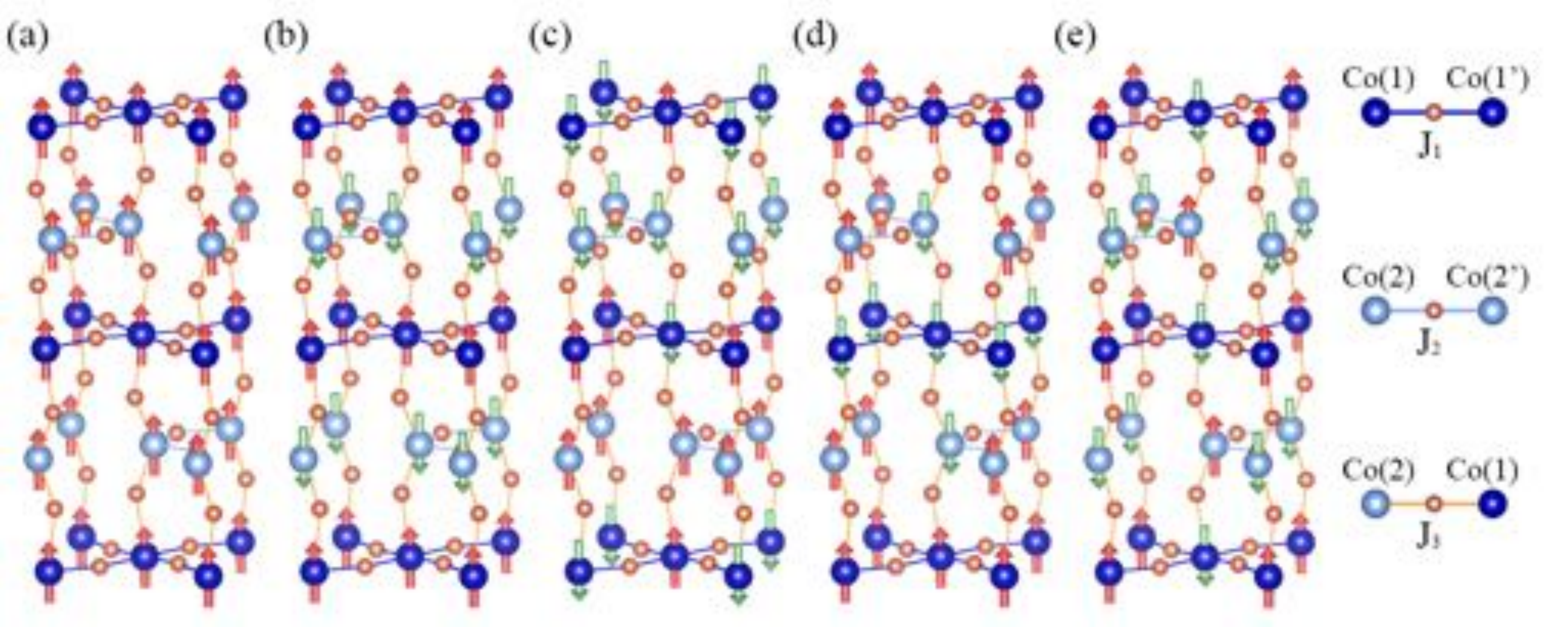}
  \caption{(Color online) Magnetic configurations considered for total energy calculations: (a) FM, (b) AFM(I), (c) AFM(II), (d) AFM(III), (e) AFM(IV). AFM(I) corresponds to A-AFM. AFM(IV) with the G-AFM ordering is the lowest energy configuration. The dark and light sphere symbols represent for Co(1) and Co(2) atoms, respectively.}
  \label{fig:ordering}
\end{figure*}

From the total energies for different spin configurations, we can extract the parameters for effective exchange interactions in BM-SCO. To calculate the parameters, we assume a Heisenberg model for the nearest-neighbor interactions between Co spins, $\mathcal{H}=E_0-\sum_{\langle ij\rangle} J_{ij} \mathbf{S}_i\cdot\mathbf{S}_j$, where $J_{ij}$ represents for the exchange interactions, $\mathbf{S}_i$ for Co spins, and $E_0$ for the reference of total energies. For the FM ground state of cubic perovskite SrCoO$_{3}$, for instance, we can assume only one parameter $J_0$ for the exchange interactions in P-SCO. From the calculated total energies of different spin configurations with $U_{\mathrm{eff}}=1.5$~eV, as shown in Fig.~\ref{fig:p-sco-u-dep}, we can determine the AFM-FM energy difference for the single bond of Co-Co spins to be 73~\textit{m}eV. From the classical energy expression for the Heisenberg model, $\Delta E_{\mathrm{AFM-FM}} = 2 J_0 S^2$ for Co $S=3/2$, the effective exchange interaction $J_0$ for P-SCO becomes 16~\textit{m}eV, which is quite consistent with the previous DFT work.\cite{Hoffmann:2015aa}

However, the mean field $T_c\approx 800$ K for P-SCO estimated from our DFT result turns out to be much higher than the experimental FM ordering temperatures of about 280~K.\cite{Long:2011aa,Bezdicka:1993aa,Balamurugan:2006aa,Kawasaki:1996} Hoffmann \emph{et al.}\cite{Hoffmann:2015aa} has extensively investigated this issue and concluded that the oxygen deficiency of the SrCoO$_{3-\delta}$ samples in the experiment might be responsible for the observed Curie temperatures. This interpretation appears to be consistent with the magnetization measurements exhibiting a large variation of $T_c$ depending on the oxygen contents in SrCoO$_{3-\delta}$.\cite{Balamurugan:2006aa} In fact, the presence of oxygen vacancy induces the increase of the Co-Co distance, as discussed in Fig.~\ref{fig:co-co-distance}, which in turn leads to the reduction of the $pd$-hybridization and the stability of FM ordering. This is another indication that the ground state of P-SCO is affected by strong electron correlations.

The magnetic exchange interactions in BM-SCO is rather complicated. According to the crystal symmetry, as shown in Fig.~\ref{fig:ordering}, there are three inequivalent $J_{ij}$ parameters: $J_1$ is for the interaction between octahedral Co atoms, i.e., Co(1) and Co(1'), $J_2$ for the interaction among Co atoms, i.e., Co(2) and Co(2'), within the tetrahedral layers and $J_3$ for the interaction between Co(1) and Co(2). We use the calculated total energies for five different spin configurations shown in Fig.~\ref{fig:ordering} and estimate the parameters for effective exchange interactions among Co spins with $S=2$ in its HS state. All the values of $J_i$'s are negative in favor of the antiferromagnetic exchange interaction between nearest neighbor Co spins, leading to the G-AFM ground state. We carried out the same analysis for the larger value of $U_{\mathrm{eff}}=3.5$ eV for comparison, but the results remain the same. Our calculation results are consistent with the previous calculation\cite{Mitra:2013aa} as well as experiment.\cite{Munoz:2008aa}

\begin{table}
	\begin{ruledtabular}
		\begin{tabular}{@{\extracolsep{\fill}} l >{\centering}m{1.1cm} >{\centering}m{1.1cm} >{\centering}m{1.1cm}}
			%\hline\hline
			\multirow{2}{*}{$U_\mathrm{eff}$ (eV)} & \multicolumn{3}{ c }{Exchange parameter (meV)}                                                                                                               \\
			                        & $J_1$ & $J_2$ & $J_3$ \tabularnewline
			\hline
			1.5                                    & -9.12 & -6.05 & -12.9 \tabularnewline
			3.5                                    & -11.6 & -6.05 & -10.4 \tabularnewline
			%\hline\hline
		\end{tabular}
	\end{ruledtabular}
	\caption{Calculated parameters for the exchange interactions in brownmillerite SrCoO$_{2.5}$ (BM-SCO). The bond configuration for each exchange interaction parameter $J_i$ is depicted in Fig.~\ref{fig:ordering}.}
	\label{tab:BMj-result}
\end{table}

\section{Intermediate Structure S\lowercase{r}C\lowercase{o}O$_{2.75}$}
\label{sec:intermediate-structure}

\subsection{Structural and electronic properties of SrCoO$_{2.75}$}
\label{sub:structure-electronic-SCO-V1}

Along the topotactic transition between P-SCO and BM-SCO, Choi \textit{et al.}\cite{Choi:2013aa} showed that SrCoO$_{3-\delta}$ ($0\le\delta\le 0.5$) exhibits a reversible lattice and electronic structure evolution according to the change of oxygen stoichiometry. They identified that the change of oxygen stoichiometry is crucial to the metal-to-insulator and ferromagnetic-to-antiferromagnetic transitions. Although the metal-to-insulator transition can be understood by extrapolating the two stable endpoints of FM metallic P-SCO ($\delta=0$) and G-AFM insulating BM-SCO ($\delta=0.5$), the correlation between magnetic ordering and metal-to-insulator transitions are not clarified yet.

To understand the multivalent nature of Co ions in SrCoO$_{3-\delta}$ ($0\le\delta\le 0.5$), along the line of the topotactic transition between P-SCO and BM-SCO, we carried out DFT calculations for the structural, electronic, and magnetic properties of SrCoO$_{2.75}$. We have already presented the structural and magnetic properties of SrCoO$_{2.75}$ with the SCO-V0 structure in Sec.~\ref{sub:electronic structure of P-SCO}. Randomly arranged oxygen vacancies in SrCoO$_{3-\delta}$ can include octahedra, tetrahedra, and pyramid inside the crystal. However, the SCO-V0 structure has simply one oxygen vacancy introduced in the $\sqrt{2}\times\sqrt{2}\times2$ supercell so that it does not have any feature connected to the brownmillerite structures.

\begin{figure}
	\centering
	\includegraphics[width=0.9\linewidth]{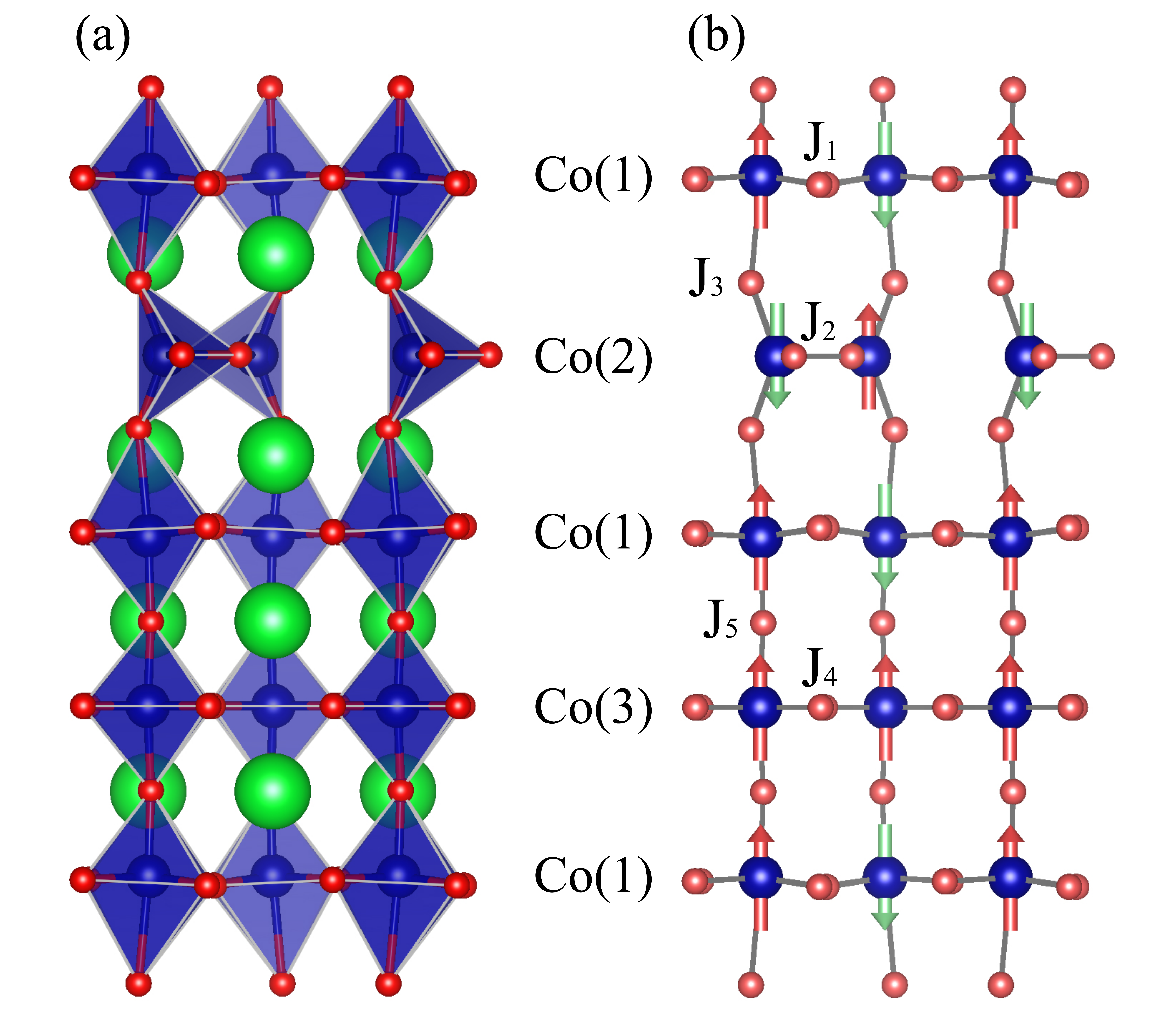}
  \caption{(Color online) (a) Relaxed unit-cell structure of the intermediate oxygen content (SCO-V1) for SrCoO$_{2.75}$ and (b) the ground-state spin configuration with the exchange interaction parameters of $J_i$'s.}
	\label{fig:v1-structure}
\end{figure}

Based on a structural model for SrCoO$_{2.75}$ suggested by experimental observation,\cite{Hao:2015aa} we constructed a structure containing only octahedra and tetrahedra. Adding two oxygen atoms to the SrCoO$_{2.5}$ BM-SCO unit cell and carrying out full-lattice relaxations, we obtained several stable structures for SrCoO$_{2.75}$. Here we present one of the typical structures for the detailed analysis of electronic and magnetic properties of SrCoO$_{2.75}$, as shown in Fig.~\ref{fig:v1-structure}(a). This structure is derived from the \textit{I2mb} unit cell of BM-SCO, and the vacancies in one of the tetrahedral layer were filled by additional oxygen atoms. From now on, we call this structure SCO-V1 to distinguish it from SCO-V0.

\begin{figure}
	\includegraphics[width=1.2\linewidth]{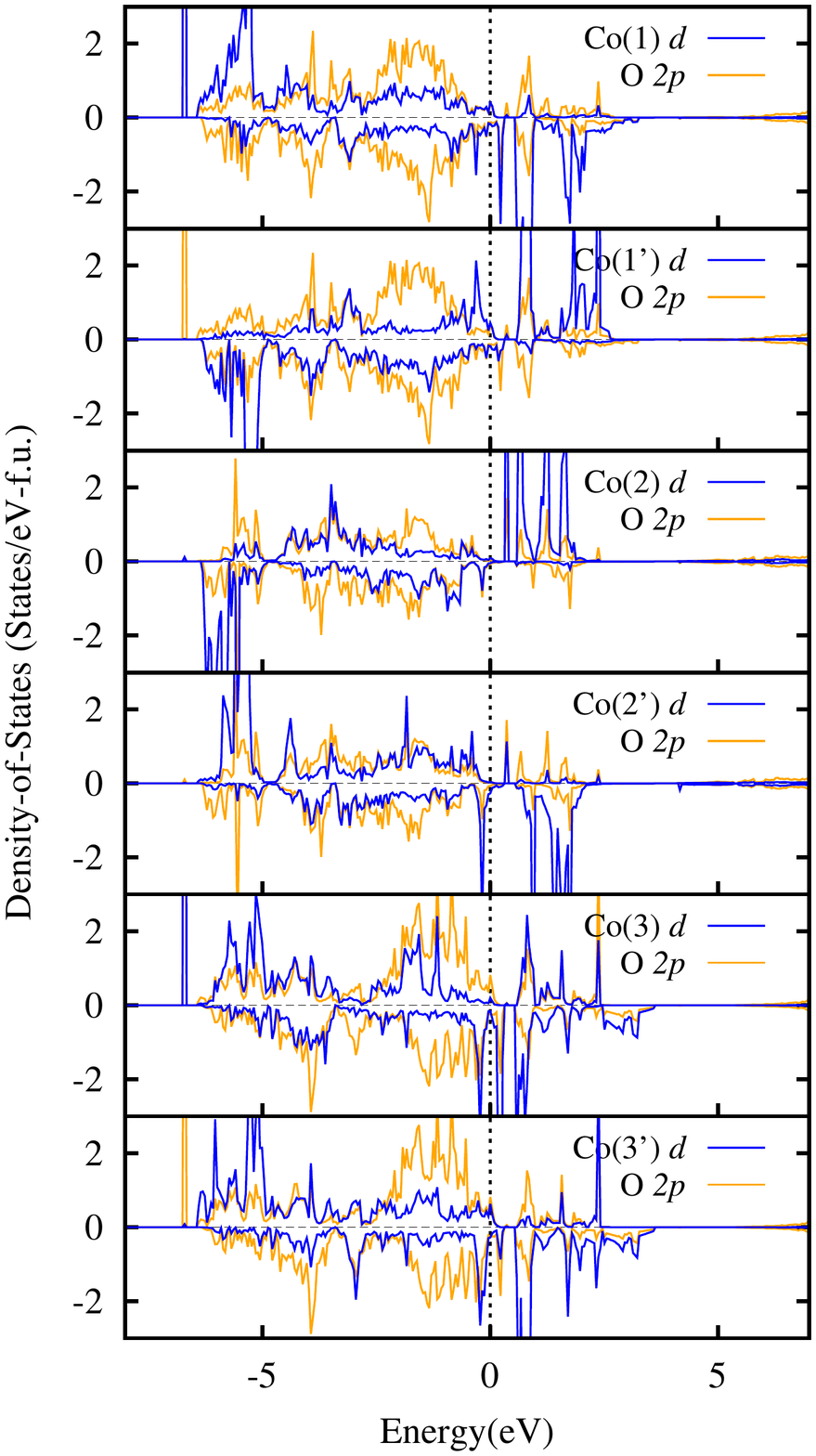}
	\caption{(Color online) Projected density-of-states for the intermediate structure (SCO-V1) of SrCoO$_{2.75}$. Each panel shows the spin-up (dark line) and spin-down (light line) density-of-states of Co $d$ and O $p$ components, respectively.}
	\label{fig:v1-sco-pdos}
\end{figure}

The SCO-V1 structure of Fig.~\ref{fig:v1-structure}(a) has mixed characteristics of both P-SCO and BM-SCO. The block of Co(1)-Co(2)-Co(1) layers resemble the BM-SCO structure, while the block of Co(3) surrounded by Co(1) layers is close to that of perovskite P-SCO. The calculated effective Co-Co distance for SCO-V1 is larger than P-SCO but smaller than BM-SCO. The effective Co-Co distances of SCO-V0 and SCO-V1 are 3.874 {\AA} and 3.932 {\AA}, respectively, which lie in order between 3.843 {\AA} of P-SCO and 3.954 {\AA} of BM-SCO. Considering the relation between Co-Co distance and magnetic ordering discussed in Fig.~\ref{fig:co-co-distance}, the effective Co-Co distance of SCO-V1 is somewhat close to that of BM-SCO. Therefore, the overall electronic structure of SCO-V1, shown in Fig.~\ref{fig:v1-sco-pdos}, is dominated by the \textit{strongly} suppressed $dp\sigma$-hybridization regardless of their local environments of P-SCO or BM-SCO. This suppression of $dp\sigma$-hybridization is quite similar to the case of BM-SCO in Fig.~\ref{fig:bm-banddos1} except that the bandgap is almost closed at the Co(3) site. These features are entirely consistent with our conjecture that the effective Co-Co distance determines the magnetic ground state by means of controlling the effective $dp\sigma$ hybridization between Co $e_{\mathrm{g}}$ and O $p\sigma$ orbitals.

The gap closing is related to the in-plane FM ordering within the Co(3) layers, which has a similar environment as P-SCO. In fact, it is noted that the A-AFM ground state is favored in the SCO-V0 structure of SrCoO$_{2.75}$. From the pDOS plots of SCO-V1 of Fig.~\ref{fig:v1-sco-pdos}, we can identify that the P-SCO block of Co(1)-Co(3)-Co(1) layers become almost metallic with the diminishing gap while the BM-SCO block of Co(1)-Co(2)-Co(1) remains insulating with a finite gap.

\subsection{Effective exchange interactions in SrCoO$_{2.75}$}
\label{sub:effective-exchange-SCO-V1}

The magnetic configurations of SCO-V1 are rather complicated but may be viewed regarding the combination of P-SCO and BM-SCO magnetic structures. As illustrated in Fig.~\ref{fig:v1-structure}(b), the ground state spin-configuration of the ground state of SCO-V1 SrCoO$_{2.75}$ is a mixture of G-AFM around Co(1) and the FM within the Co(3) layer. We can assign a similar set of exchange parameters $J_1$, $J_2$ and $J_3$ between Co(1) and Co(2) spins as defined in the BM-SCO SrCoO$_{2.5}$. As for the spins surrounding Co(3), which resembles the P-SCO environment, additional parameters of $J_4$ and $J_5$ are introduced. $J_4$ represents for the intra-layer interaction within the Co(3) layer, similar to $J_1$, while $J_5$ represents for the inter-layer interaction between Co(1) and Co(3). Following the same procedures discussed in Sec.~\ref{sub:effective-exchange-BM-SCO}, we determine the $J_i$'s from the total energies for several spin configurations. The parameters of effective exchange interactions in SCO-V1 SrCoO$_{2.75}$ are listed in Table~\ref{tab:A1j-result}. This result indicates that a strong ferromagnetic interaction develops within the Co(3) layer, which may be the reflection of the A-AFM ordering in the SCO-V0 structure, while the antiferromagnetic interaction dominates in other regions where the oxygen vacancy ordering of BM-SCO dominates.

\begin{table}
	%\centering
	\begin{ruledtabular}
		\begin{tabular}{l >{\centering}m{1.2cm} >{\centering}m{1.2cm} >{\centering}m{1.2cm} >{\centering}m{1.2cm} >{\centering}m{1.2cm}}
			%\hline\hline
			\multirow{2}{*}{$U_\mathrm{eff}$ (eV)} & \multicolumn{5}{c}{Exchange parameters (meV)} \tabularnewline
			                                       & $J_1$                                                         & $J_2$ & $J_3$ & $J_4$ & $J_5$\tabularnewline \hline
			1.5                                    & -9.57                                                         & -16.6 & -11.1 & 7.88  & 9.39 \tabularnewline
			3.5                                    & -6.21                                                         & -16.3 & -13.2 & 8.56  & 7.42 \tabularnewline
			%\hline\hline
		\end{tabular}
	\end{ruledtabular}
	\caption{The calculated exchange parameters of the intermediate structure (SCO-V1) of SrCoO$_{2.75}$. The bond configuration for each exchange interaction parameter $J_i$ is depicted in Fig.~\ref{fig:v1-structure}(b). It is noted that the positive values of $J_4$ and $J_5$ support the ferromagnetic ordering within the Co(3) layers.}
	\label{tab:A1j-result}
\end{table}

\section{Discussion}
\label{sec: discussion}

\begin{figure}
	\includegraphics[width=0.9\linewidth]{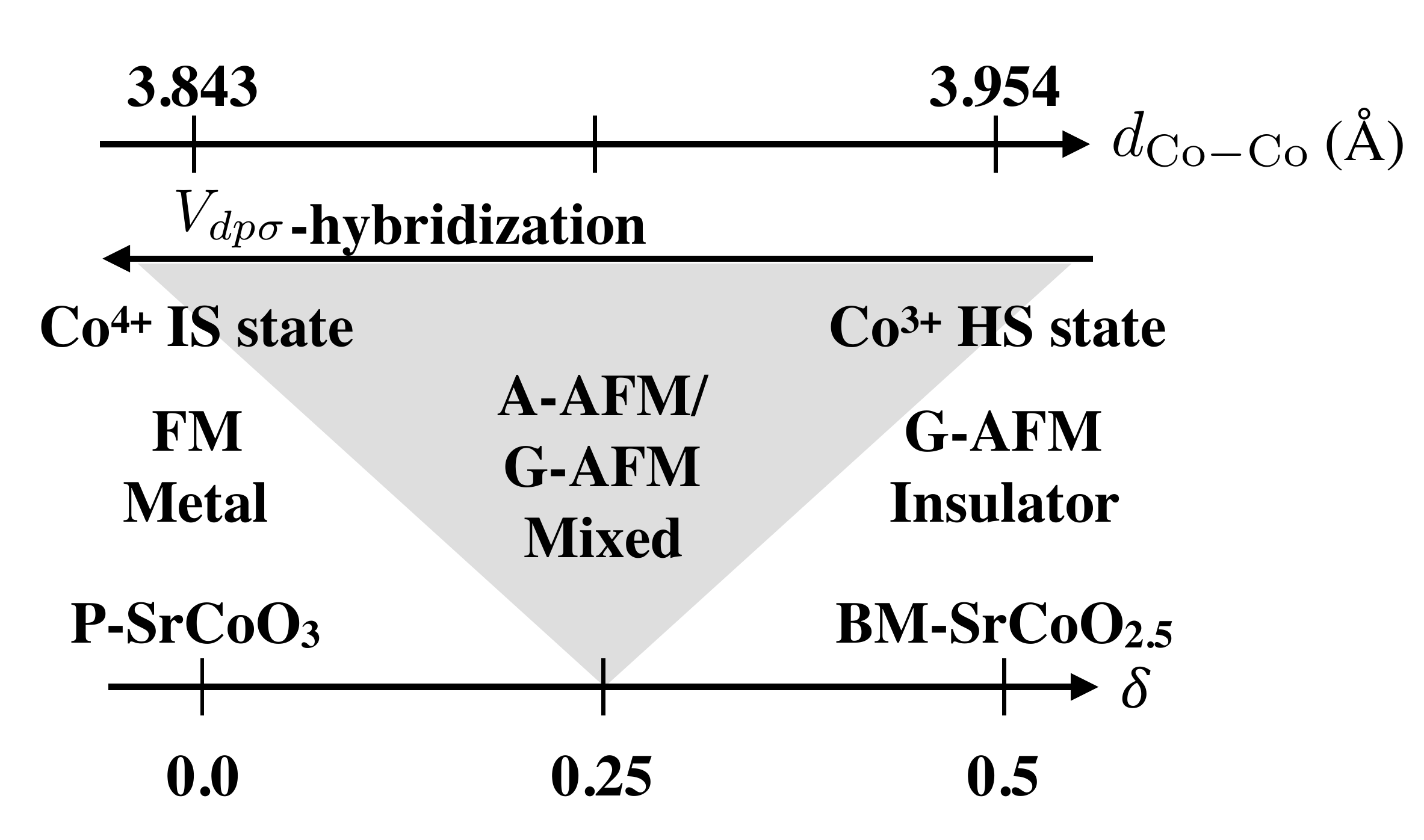}
	\caption{Schematic phase diagram of SrCoO$_{3-\delta}$ along the line of topotactic transition between FM perovskite SrCoO$_{3}$ and G-AFM brownmillerite SrCoO$_{2.5}$. The increase of lattice constant, i.e., the effective Co-Co distance, $d_{\mathrm{Co-Co}}$, with the increase of oxygen vacancy, i.e., $\delta$, reduces the $V_{dp\sigma}$-hybridization and drives the IS-to-HS spin transition as well as the metal-to-insulator transition from P-SCO to BM-SCO.}
	\label{fig:phase-diagram}
\end{figure}

In this work, we have investigated the structural, electronic, and magnetic properties of SrCoO$_{3-\delta}$ ($\delta=0, 0.25, 0.5$) by carrying out first-principles calculations by using density-functional theory within the GGA+$U$ method. To understand the multivalent nature of Co ions in oxygen-deficient SrCoO$_{3-\delta}$ along the line of the topotactic transition between P-SCO and BM-SCO, we examine the $U_{\mathrm{eff}}$-dependent magnetic structures and address the issues of the proximity of the FM-metallic P-SCO to other AFM states. As illustrated in Fig.~\ref{fig:phase-diagram}, the results provide an insight into the relationship between the Co-Co distances and the magnetic couplings where the change of the Co spin state is driven by the change of $pd$-hybridization and the effective Co-Co distance. We also analyze the effect of oxygen vacancy on the lattice volume and the effective Co-Co distances. The increase of the lattice volume obtained in our calculations is consistent with the previous theoretical work, \cite{Cazorla:2017aa} which supports the formation of oxygen vacancy mediated by the strain control.\cite{Jeen:2013ab,Petrie:2016aa,Hu:2017aa} The increase of effective Co-Co distances leads to the reduction of the effective $pd$-hybridization. Since the $pd$-hybridization between Co and O atoms affects the crystal field strength, in turn, the Co spin state is strongly influenced by the change of Co-Co distances.

It appears that the spin-state transition accompanying the topotactic transition is driven by the change of oxygen content in SrCoO$_{3-\delta}$, but the real mechanism behind such transition is the suppression of the Co $d$-O $p$ hybridization triggered by the oxygen vacancy and, in turn, the increase of lattice volume. It is confirmed that the electronic and magnetic structure of P-SCO is in proximity to the IS and HS states with either FM or AFM ordering depending on the strength of on-site Coulomb interactions and oxygen vacancy as well. Thus, it is no wonder that SrCoO$_{3-\delta}$ in such a critical state exhibits a strong spin-phonon coupling effect.\cite{Lee:2011ab} Even the epitaxial strain can induce phase transitions from FM-metal to AFM-insulator with ferroelectricity.\cite{Lee:2011aa} Nevertheless, it is interesting to note that the BM-SCO with the \textit{I2mb} structure, which is found to be energetically the most stable, exhibits ferroelectricity via the ordering of CoO$_{4}$ tetrahedra and the polar structure of \textit{I2mb} may be stabilized by the large activation barrier among different structural configurations. Maybe further experimental studies are necessary to prove the existence of the ferroelectricity in brownmillerite SrCoO$_{2.5}$.

Besides the issue of ferroelectricity, the magnetic ordering of Co spins in the oxygen-deficient SrCoO$_{3-\delta}$ is crucial to its electronic and magnetic properties. We observe that the FM spin configurations of SrCoO$_{2.75}$ are critical at the metal-to-insulator boundary, while the electronic structure of P-SCO is FM-metallic and that of BM-SCO ($\delta=0.5$) with the G-type antiferromagnetic (G-AFM) ordering is insulating. Since the local spin configuration is also crucial to the electronic structure properties, the overall electronic structure can be complicated depending on the spin configurations. Therefore, we conclude that the magnetic ground states of SrCoO$_{3-\delta}$ are closely tied up with the effective Co-Co distances, which depend on the contents of oxygen vacancy. Near the metal-to-insulator boundary, for example, close to SrCoO$_{2.75}$, the local spin configuration is shown to be critical to the formation of a bandgap. Since the structural configuration can be complicated, the overall state of electronic conduction may be determined by the conducting path formed primarily by the FM ordered layers as described in Fig.~\ref{fig:v1-structure}. The conducting channels in the mixed spin configurations of SrCoO$_{3-\delta}$ can be attributed to the origin of the observed resistance switching in epitaxial SrCoO$_{3-
\delta}$ thin films.\cite{Tambunan:2014aa}

%\section*{Acknowledgements}
\begin{acknowledgments}
  We gratefully acknowledge Tamio Oguchi and Woo Seok Choi for valuable discussions.	This work was supported by the National Research Foundation of Korea (NRF) (no. 2017R1A2B4007100). JY gratefully acknowledges the support and hospitality provided by the Max Planck Institute for the Physics of Complex Systems, where this work was completed during his visit to the institute.
\end{acknowledgments}

%\section*{References}
%\bibliographystyle{iopart-num}
\bibliography{srcoo3}

\end{document}